\documentclass[aps,prd,twocolumn,10pt]{revtex4-2}

\usepackage{amsmath,amssymb}
\usepackage{booktabs}
\usepackage{graphicx}
\usepackage[colorlinks=true,allcolors=blue]{hyperref}
\usepackage[normalem]{ulem}
\usepackage{xcolor}

\usepackage{xcolor}
\usepackage{soul}

\setstcolor{red}  
\begin{document}

\title{Probing Axion Like Particle-Proton Interactions through the Interplay of Dark Matter and Cosmic Rays in Large-Scale Structure}

\author{Wen-Qing Guo}
\email{wetingguo@gmail.com}
\affiliation{Department of Physics, Stellenbosch University, Matieland 7602, South Africa}

\begin{abstract}
Axion-like particles (ALPs) provide a well-motivated dark matter candidate whose interactions with Standard Model particles may generate observable cosmological signatures. We investigate the diffuse $\gamma$-ray emission produced by the scattering process between ALP dark matter and cosmic-ray (CR) protons inside dark matter halos, and explore its detectability through the cross-correlation between the unresolved $\gamma$-ray background (UGRB) measured by \textit{Fermi} Large Area Telescope (LAT) and the Two Micron All Sky Survey Redshift Survey (2MRS) galaxy catalog in the local Universe ($z\leq0.1$). To model the cosmological $\mathrm{GeV}$--$\mathrm{PeV}$ CR proton population, we develop a phenomenological bottom-up prescription that links CR injection to the star formation rate through the supernova rate, while the steady-state CR density is determined by diffusive escape. Our forecasts indicate that UGRB-galaxy cross-correlation measurements can probe previously unexplored regions of the ALP parameter space over a broad range of ALP masses. More generally, this work presents a general framework for cosmological observables generated by interactions between dark matter and an additional astrophysical field, opening a new avenue for indirect dark matter searches using large-scale structure observations.
\end{abstract}

\maketitle

\section{Introduction}

The existence of dark matter (DM) is firmly established by a wealth of astrophysical and cosmological observations. Within the standard $\Lambda$CDM cosmological model, measurements of the cosmic microwave background (CMB) indicate that approximately $26\%$ of the total energy density of the Universe is composed of non-baryonic DM \cite{Planck:2018vyg}. Although its gravitational effects have been extensively confirmed over a wide range of scales, the particle nature of DM remains one of the most fundamental open questions in modern physics.

Among the numerous candidates proposed beyond the Standard Model (SM), axion-like particles (ALPs) have attracted considerable attention because they arise naturally in many extensions of the SM, particularly string-inspired theories \cite{Marsh:2015xka, Irastorza:2018dyq, Co:2020xlh}. Unlike the QCD axion, ALPs are not tied to the Peccei-Quinn solution to the strong CP problem, allowing their masses and couplings to span a broad and largely unconstrained parameter space \cite{Irastorza:2021tdu}. Depending on their production history and cosmological evolution, such as through non-thermal production mechanisms that allow relatively heavy ALPs to behave as cold DM, they can constitute all or part of the cosmological DM abundance. Consequently, extensive experimental and observational efforts have been devoted to probing ALPs through laboratory searches \cite{Bahre:2013ywa, ADMX:2019uok, Mostepanenko:2020lqe}, stellar environments \cite{CAST:2017uph, Hook:2018iia, Bhusal:2020bvx, Buschmann:2021juv, Foster:2022fxn, Lella:2023bfb, Li:2025zgp}, dwarf spheroidal galaxies \cite{Caputo:2018ljp, Guo:2024oqo}, and other astrophysical systems \cite{Fermi-LAT:2016nkz, Caputo:2018vmy, Xia:2019yud, Li:2020pcn, Guo:2026cre}.

Recently, increasing attention has been devoted to $\gamma$-ray signatures induced by interactions between ALP dark matter and high-energy cosmic rays (CRs) \cite{Dent:2020qev, Goncalves:2025nij, Goncalves:2026ean}. In particular, the scattering process $a+p\rightarrow p+\gamma$, mediated by the ALP-proton coupling, enables non-relativistic ALPs in the $\text{keV}$-to-$\text{GeV}$ mass range to produce observable gamma rays in environments permeated by relativistic CR protons. Assuming the ALP-photon coupling is sufficiently suppressed such that the ALP lifetime substantially exceeds the age of the Universe, this scattering channel offers a distinct window into heavy ALP scenarios. Owing to the small interaction cross-section, this process generates a faint but potentially detectable $\gamma$-ray signal. Existing studies have primarily focused on $\gamma$-ray emission from local astrophysical systems. While these investigations have provided valuable constraints on the ALP-proton coupling, they probe relatively small angular scales, constrained by the local environments of specific DM halos.

From a cosmological perspective, dark matter halos naturally provide environments in which both ingredients required for this interaction coexist. ALP dark matter permeates every halo, while CR protons are continuously accelerated by astrophysical activities associated with galaxies residing within the same halos. Consequently, ALP-CR proton interactions are expected to occur throughout the halo population across cosmic time, producing diffuse $\gamma$-ray emission on cosmological scales. Since the overwhelming majority of these halos cannot be individually resolved by current $\gamma$-ray observations, their cumulative emission contributes naturally to the unresolved $\gamma$-ray background (UGRB), which has been accurately measured over a broad energy range by the \textit{Fermi} Large Area Telescope (LAT) \cite{Fermi-LAT:2012pez, Fermi-LAT:2014ryh, Fermi-LAT:2018udj}.

The UGRB is generally believed to be dominated by unresolved astrophysical source populations, including BL Lacertae objects (BL Lac), flat spectrum radio quasars (FSRQ), misaligned active galactic nuclei (mAGN), and star-forming galaxies (SFGs) \cite{Ando:2009nk, Xia:2011ax, DiMauro:2014wha, Xia:2015wka, Hooper:2016gjy, Cuoco:2017bpv}. Nevertheless, the relative contributions of these populations remain uncertain, leaving room for additional components, such as $\gamma$-ray emission induced by dark matter \cite{Ando:2005xg, Fornengo:2013rga, Fornasa:2016ohl, Pinetti:2019ztr, Zhou:2024cld}. Measurements of the UGRB intensity alone provide only limited information for disentangling these different contributions. A more powerful approach is to exploit their spatial correlations with independent tracers of the large-scale structure. In particular, cross-correlating the UGRB with galaxy catalogs \cite{Regis:2015zka, Ammazzalorso:2018evf, Pinetti:2025hgd} or cosmic shear measurements \cite{Camera:2012cj, Camera:2014rja, Zhang:2026ysp} provides valuable tomographic information that significantly improves the ability to distinguish different $\gamma$-ray source populations.

In this work, we investigate the cosmological $\gamma$-ray signature produced by ALP-CR proton interactions and assess its detectability through the cross-correlation between the \textit{Fermi}-LAT UGRB and the Two Micron All Sky Survey Redshift Survey (2MRS) galaxy catalog. Extending the local analysis of Ref.~\cite{Goncalves:2026ean} to cosmological scales, we compute the $\gamma$-ray emission generated by ALP-CR proton scattering within individual dark matter halos and evaluate its contribution to the UGRB angular power spectrum (APS) using the halo model. A key ingredient of this calculation is the $\mathrm{GeV}$--$\mathrm{PeV}$ CR proton population inside dark matter halos. Since this quantity cannot be directly inferred for the cosmological halo population, we develop a physically motivated bottom-up model based on halo properties and empirical astrophysical scaling relations to predict the CR distribution self-consistently. This framework enables us to forecast the sensitivity of UGRB-galaxy cross-correlation measurements to the ALP-proton coupling over a broad ALP mass range.

This paper is organized as follows. In Sec.~\ref{Sec:Theore}, we delineate the theoretical framework for the angular power spectrum. The forecast methodology and discussion are described in Sec.~\ref{Sec:ResultDiscussion}, followed by our conclusions in Sec.~\ref{Sec:Conclusion}.

Throughout this paper, we adopt a spatially flat $\Lambda$CDM cosmology with the \textit{Planck} 2018 cosmological parameters: $H_{0}=67.32~{\rm km\,s^{-1}Mpc^{-1}}$, $\Omega_{\rm c}h^{2}=0.1201$, and $\Omega_{\rm b}h^{2}=0.02238$ \cite{Planck:2018vyg}.

\section{The cross-correlation signal}\label{Sec:Theore}
\subsection{Angular power spectrum}
We characterize the cross-correlation between two cosmological observables, denoted as $i$ and $j$, using the cross angular power spectrum. Under the Limber approximation, the cross APS $C_{\ell}^{ij}$ can be expressed as a line-of-sight integral \cite{Fornengo:2013rga}
\begin{equation}\label{eq:APS}
C_{\ell}^{ij} = \int \frac{\mathrm{d}\chi}{\chi^{2}} W_i(\chi) W_j(\chi) P_{ij}\left(z, k=\frac{\ell+1/2}{\chi}\right),
\end{equation}
where $\ell$ represents the angular multipole, $\chi(z)$ is the comoving distance determined by $\mathrm{d}\chi = c\,\mathrm{d}z/H(z)$, and $W(\chi)$ denotes the window function describing the redshift dependence of each observable. The Hubble expansion rate $H(z)$ in a flat $\Lambda$CDM background evolves as
\begin{equation}
H(z) = H_0 \sqrt{\Omega_{\rm m}(1+z)^3 + \Omega_\Lambda},
\end{equation}
where $\Omega_\Lambda = 1 - \Omega_{\rm m}$, and $\Omega_{\rm m} = 0.3158$. Throughout this work, comoving distance $\chi$ and redshift $z$ are used interchangeably.

The term $P_{ij}(z,k)$ in Eq.~\eqref{eq:APS} represents the three-dimensional cross-power spectrum of the fluctuation fields associated with the two observables. Within the halo-model framework \cite{Cooray:2002dia}, $P_{ij}(z,k)$ is decomposed into one-halo (1h) and two-halo (2h) contributions
\begin{equation}
P_{ij}(z,k) = P_{ij}^{1\mathrm{h}}(z,k) + P_{ij}^{2\mathrm{h}}(z,k).
\end{equation}

The one-halo term represents the correlation between two tracers residing within the same host dark matter halo, dominating on small spatial scales. Conversely, the two-halo term describes the correlations between tracers hosted by two distinct halos, which dominates on large, linear scales. Their explicit formulations are given by
\begin{align}
P^{1\mathrm{h}}_{ij}(z,k) &= \int \mathrm{d}M \frac{\mathrm{d}n}{\mathrm{d}M}(z,M) \frac{\tilde g_i^{*}(k|z,M)}{\langle g_i\rangle(z)} \frac{\tilde g_j(k|z,M)}{\langle g_j\rangle(z)}, \\
P^{2\mathrm{h}}_{ij}(z,k) &= \left[ \int \mathrm{d}M_1 \frac{\mathrm{d}n}{\mathrm{d}M_1} b_{\rm h}(z,M_1) \frac{\tilde g_i^{*}(k|z,M_1)}{\langle g_i\rangle(z)} \right] \nonumber \\
&\quad \times \left[ \int \mathrm{d}M_2 \frac{\mathrm{d}n}{\mathrm{d}M_2} b_{\rm h}(z,M_2) \frac{\tilde g_j(k|z,M_2)}{\langle g_j\rangle(z)} \right] P_{\rm lin}(z,k),
\end{align}
where $M$ denotes the host halo mass, and the integrations are performed over a wide mass range spanning from $M_{\rm min} = 10^{-6}\,\mathrm{M_\odot}$ to $M_{\rm max} = 10^{18}\,\mathrm{M_\odot}$. The quantities $\mathrm{d}n/\mathrm{d}M$ and $b_{\rm h}$ represent the halo mass function (HMF) and the halo bias, respectively, both parameterized using the Sheth-Tormen prescriptions \cite{Sheth:1999su, Sheth:1999mn}. The function $\tilde g(k|z,M)$ corresponds to the three-dimensional Fourier transform of the spatial profile $g(r|z,M)$ of a given tracer within a halo of mass $M$, with the superscript ``$*$'' denoting complex conjugation. The linear matter power spectrum $P_{\rm lin}(z,k)$ is computed using the public code {\tt CAMB}\footnote{\url{https://camb.readthedocs.io/en/latest/}}.

In this work, we focus on the cross-correlation between 2MRS galaxy catalog and the UGRB measured by the \textit{Fermi}-LAT. The specific window functions, spatial profiles, and individual halo-model ingredients for these two distinct tracers are detailed in the following subsections.

\begin{figure*}[t]
    \centering
    \includegraphics[width=0.48\linewidth]{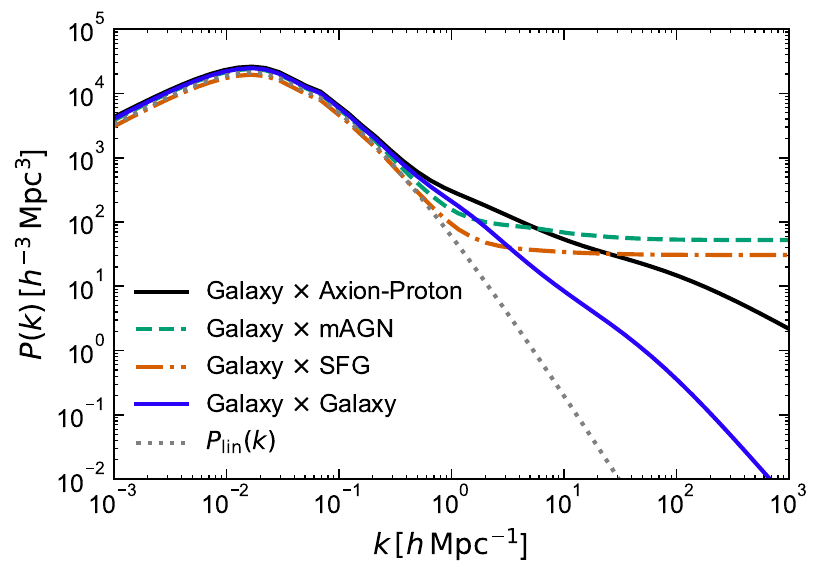}
    \includegraphics[width=0.48\linewidth]{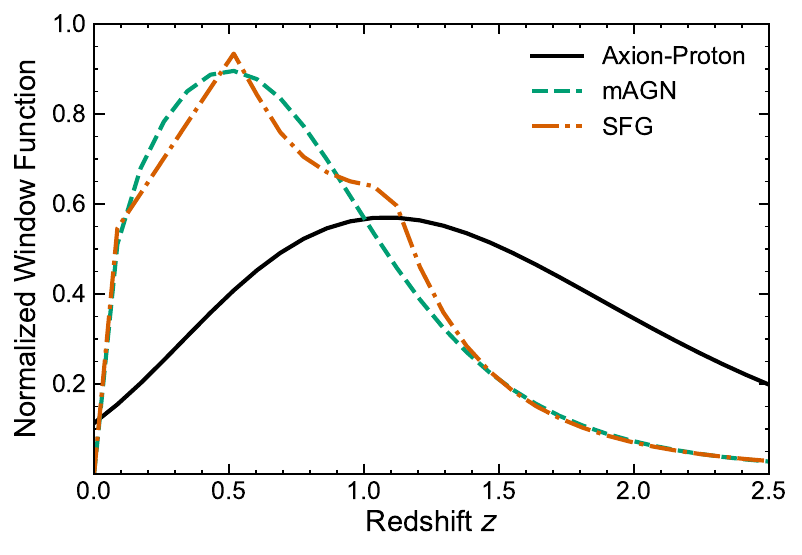}
    \caption{\textit{Left}: Three-dimensional power spectra at $z=0.05$. The solid blue line shows the galaxy auto-power spectrum, while the solid black, dashed green, and dash-dotted brown lines show the galaxy-$\gamma$-ray cross-power spectra for ALP-CR proton-induced emission, unresolved mAGNs, and unresolved SFGs, respectively, in the $(1.0,\,1.7)\,\mathrm{GeV}$ energy bin. The linear matter power spectrum at the same redshift is shown by the gray dotted line. \textit{Right}: Normalized window functions of the different $\gamma$-ray components as a function of redshift in the $(1.0,\,1.7)\,\mathrm{GeV}$ energy bin. The solid black, dashed green, and dash-dotted brown curves correspond to ALP-CR proton-induced emission with $m_a=1\,\mathrm{MeV}$, unresolved mAGNs, and unresolved SFGs, respectively. 
     }
    \label{fig:WindowF_and_PowerS}
\end{figure*}

\subsection{Galaxy catalog}
To trace the large-scale matter distribution in the local Universe, we adopt the Two Micron All Sky Survey Redshift Survey \cite{2012ApJS..199...26H}, which provides an exceptionally complete spectroscopic galaxy catalog at low redshifts. The specific 2MRS sample contains spectroscopic redshifts for $N_{\rm g} = 43,182$ galaxies extending out to $z \simeq 0.1$, featuring an effective sky coverage of $f_{\rm 2MRS} = 0.877$ after applying standard masking \cite{2006AJ....131.1163S, Ando:2017wff}. Following Ref.~\cite{Ando:2017wff}, the normalized window function of the 2MRS galaxies is parameterized as
\begin{equation}
\frac{\mathrm{d}N_{\rm 2MRS}}{\mathrm{d}z} = \frac{\beta}{z_0\Gamma[(m+1)/\beta]} \left( \frac{z}{z_0} \right)^m \exp \left[ - \left( \frac{z}{z_0} \right)^\beta \right],
\end{equation}
where the empirical best-fit parameters are given by $\beta = 1.64$, $z_0 = 0.0266$, and $m = 1.31$, and $\Gamma(x)$ denotes the Gamma function. The resulting galaxy window function incorporated into the angular power spectrum in Eq.~\eqref{eq:APS} is defined as
\begin{equation}
W_{\rm g}(\chi) = \frac{\mathrm{d}N_{\rm 2MRS}}{\mathrm{d}z} \frac{H(z)}{c}.
\end{equation}

To establish the connection between galaxies and the underlying dark matter distribution, we implement the Halo Occupation Distribution (HOD) framework \cite{Zheng:2004id, Ando:2014aoa, Ammazzalorso:2018evf}. Within this prescription, the galaxy population residing inside a host halo of mass $M$ is deterministically separated into a single central galaxy fixed at the halo center, and a swarm of satellite galaxies that spatially trace the smooth halo profile. Consequently, the galaxy number density field within a halo can be formulated as
\begin{equation}
\begin{aligned}
g_{\rm g}(\mathbf{x}-\mathbf{x}'|z,M)=&\,N_{\rm cen}(M)\delta_{\rm D}^{(3)}(\mathbf{x}-\mathbf{x}')\\
&+N_{\rm sat}(M)\frac{\rho_{\rm DM}(\mathbf{x}-\mathbf{x}'|z,M)}{M},
\end{aligned}
\end{equation}
where $\delta_{\rm D}^{(3)}$ represents the three-dimensional Dirac delta function. We model the halo profile by the standard Navarro-Frenk-White (NFW) profile
\begin{equation}
\rho_{\rm DM}(r) = \frac{\rho_{\rm s}}{(r/r_{\rm s})(1+r/r_{\rm s})^2},
\end{equation}
where the scale radius is related to the virial radius by $r_{\rm s}(z, M) = R_{\rm vir}(z, M)/c_{\rm vir}(z, M)$, with $R_{\rm vir}(z, M)$ and $c_{\rm vir}(z, M)$ denoting the halo virial radius and the concentration parameter, respectively \cite{Coe:2010xg}. The characteristic scale density $\rho_{\rm s}$ is given by
\begin{equation}
\rho_{\rm s} = \frac{M}{4\pi r_{\rm s}^3} \left[ \ln(1+c_{\rm vir}) - \frac{c_{\rm vir}}{1+c_{\rm vir}} \right]^{-1}.
\end{equation}

The mean occupation numbers for the central and satellite galaxy components are parameterized as functions of host halo mass
\begin{align}
N_{\rm cen}(M) &= \frac{1}{2} \left[ 1 + \mathrm{erf} \left( \frac{\log_{10}M - \log_{10}M_{\rm min}}{\sigma_{\log_{10} M}} \right) \right], \\
N_{\rm sat}(M) &= \begin{cases} \left( \frac{M-M_0}{M_1} \right)^\alpha, & M \ge M_0, \\ 0, & M < M_0, \end{cases}
\end{align}
where $\mathrm{erf}$ represents the error function. We adopt the empirical 2MRS best-fit HOD parameters with $\log_{10}(M_{\rm min}/\mathrm{M}_\odot) = 11.68$, $\sigma_{\log_{10}M} = 0.15$, $\log_{10}(M_{0}/\mathrm{M}_\odot) = 11.86$, $\log_{10}(M_{1}/\mathrm{M}_\odot) = 13.0$, and $\alpha = 1.02$ \cite{Ando:2014aoa}.

\begin{table*}[t]
\centering
\setlength{\tabcolsep}{10pt}
\caption{Adopted \textit{Fermi}-LAT observational specifications used in this work, following Ref.~\cite{Fermi-LAT:2018udj}. For each energy bin, we list the energy range, the photon Poisson noise $C_{\rm N}^{\gamma}$, and the effective sky fraction after masking the Galactic plane and resolved $\gamma$-ray sources.}
\label{tab:FermiLAT_specifications}
\begin{tabular}{lcccc}
\toprule
Energy Bin &
$E_{\rm min}$ [GeV] &
$E_{\rm max}$ [GeV] &
$C_{\rm N}^{\gamma}$ [cm$^{-4}$\,s$^{-2}$\,sr$^{-1}$] &
$f_{\rm Fermi}$ \\
\midrule
1  & 0.5   & 1.0    & $1.056\times10^{-17}$ & 0.199 \\
2  & 1.0   & 1.7    & $3.548\times10^{-18}$ & 0.250 \\
3  & 1.7   & 2.8    & $1.375\times10^{-18}$ & 0.443 \\
4  & 2.8   & 4.8    & $8.324\times10^{-19}$ & 0.511 \\
5  & 4.8   & 8.3    & $3.904\times10^{-19}$ & 0.564 \\
6  & 8.3   & 14.5   & $1.768\times10^{-19}$ & 0.586 \\
7  & 14.5  & 22.9   & $6.899\times10^{-20}$ & 0.586 \\
8  & 22.9  & 39.8   & $3.895\times10^{-20}$ & 0.586 \\
9  & 39.8  & 69.2   & $1.576\times10^{-20}$ & 0.586 \\
10 & 69.2  & 120.2  & $6.205\times10^{-21}$ & 0.586 \\
11 & 120.2 & 331.1  & $3.287\times10^{-21}$ & 0.586 \\
12 & 331.1 & 1000.0 & $5.094\times10^{-22}$ & 0.586 \\
\bottomrule
\end{tabular}
\end{table*}

By averaging over the halo mass function, the mean galaxy number density is obtained via
\begin{equation}
\langle g_{\rm g}\rangle(z) = \int \mathrm{d}M \frac{\mathrm{d}n}{\mathrm{d}M}(z,M) \int \mathrm{d}^3\mathbf{x}\, g_{\rm g}(\mathbf{x}-\mathbf{x}'|z,M).
\end{equation}
Accordingly, the three-dimensional Fourier transform of the galaxy profile utilized within our halo-model integrals yields
\begin{equation}
\tilde g_{\rm g}(k|z,M) = N_{\rm cen}(M) + \frac{N_{\rm sat}(M)}{M} \tilde\rho_{\rm DM}(k|z,M),
\end{equation}
where $\tilde\rho_{\rm DM}(k|z,M)$ denotes the Fourier transform of the NFW
density profile. For completeness, the effective linear galaxy bias $b_{\rm g}(z)$ is defined as
\begin{equation}
b_{\rm g}(z)
=
\int
{\rm d}M
\frac{{\rm d}n}{{\rm d}M}
b_{\rm h}(z,M)
\frac{
N_{\rm g}(M)}
{\langle g_{\rm g}\rangle(z)},
\end{equation}
where $
N_{\rm g}(M)
=
N_{\rm cen}(M)
+
N_{\rm sat}(M)$.

Utilizing the HOD formulations detailed above, we compute the auto-power spectrum of the 2MRS galaxy sample within the halo model. The resulting galaxy power spectrum evaluated at a representative redshift of $z=0.05$ is shown as the solid blue curve in the left panel of Fig.~\ref{fig:WindowF_and_PowerS}, with the corresponding linear matter power spectrum indicated by the dotted gray line.

\subsection{$\gamma$-ray emission}

To forecast the sensitivity of the UGRB-galaxy cross-correlation to the ALP-CR proton-induced signature, we adopt the observational specifications of the \textit{Fermi}-LAT, following the analysis in Ref.~\cite{Fermi-LAT:2018udj}. The UGRB dataset is partitioned into 12 energy bins spanning from $0.5~\mathrm{GeV}$ to $1~\mathrm{TeV}$. For each distinct energy bin, we incorporate the corresponding sky fraction available after masking the Galactic plane and resolved sources, alongside the photon Poisson noise reported by the \textit{Fermi}-LAT collaboration \cite{Fermi-LAT:2018udj} (see Table~\ref{tab:FermiLAT_specifications} for a summary of the instrumental parameters).

The total UGRB intensity $I_{\gamma}^{\rm tot}$ modeled in this work is expressed as the linear superposition of the unresolved conventional astrophysical background and the ALP-CR proton-induced component
\begin{equation}
I_{\gamma}^{\rm tot} = I_{\gamma}^{\rm astro} + I_{\gamma}^{\rm ALP},
\end{equation}
where $I_{\gamma}^{\rm astro}$ represents the cumulative contribution from unresolved astrophysical sources, while $I_{\gamma}^{\rm ALP}$ denotes the exotic diffuse flux arising from the $a+p\rightarrow p+\gamma$ scattering channel. Given that this study focuses on the relatively local Universe ($z \lesssim 0.1$), we restrict our astrophysical baseline to the two dominant low-redshift source classes, namely mAGNs and SFGs. These populations have been demonstrated to constitute the overwhelming majority of the astrophysical UGRB anisotropy within this redshift regime \cite{Pinetti:2019ztr, Blanco:2021icw, Zhou:2024cld}.

The formal mathematical description for unresolved astrophysical emitters is uniform across different source classes. We therefore establish the general framework here before specializing to the individual physical models for mAGNs and SFGs.

For a given unresolved astrophysical population, the distribution of sources is statistically characterized by the $\gamma$-ray luminosity function (GLF), defined as
\begin{equation}
\phi_{\gamma}(z, L_{\gamma}) = \frac{\mathrm{d}^{2}n}{\mathrm{d}V\,\mathrm{d}\ln L_{\gamma}},
\end{equation}
which represents the comoving number density of sources per unit logarithmic luminosity interval. The corresponding ensemble-averaged $\gamma$-ray luminosity density is thus given by
\begin{equation}
\langle g_{\star}\rangle(z) = \int_{L_{\rm min}}^{L_{\rm th}(z)} \mathrm{d}L_{\gamma}\, \phi_{\gamma}(z, L_{\gamma}),
\end{equation}
where the subscript ``$\star$'' denotes astrophysical sources, and $L_{\rm th}(z)$ is the redshift-dependent detection threshold of the instrument. Sources exhibiting luminosities greater than $L_{\rm th}(z)$ are resolved and excised from the data, meaning only sub-threshold sources contribute to the unresolved signal. The derivation of $L_{\rm th}(z)$ from the nominal \textit{Fermi}-LAT flux sensitivity is detailed in Appendix~\ref{app:details}.

The window function for astrophysical sources, evaluated at the observed photon energy $E_{\gamma}$, is given by \cite{Ando:2013ff}
\begin{equation}
\begin{aligned}
W_\star(E_{\gamma},\chi) = & \frac{d_{\rm L}^2(z)}{(1+z)^2} \int_{L_{\rm min}}^{L_{\rm th}(z)} \mathrm{d}\ln L_{\gamma} \, \phi_{\gamma}(z, L_{\gamma}) \\
& \times \frac{\mathrm{d}F}{\mathrm{d}E_{\gamma}}(E_{\gamma},z,L_{\gamma}) e^{-\tau([1+z]E_{\gamma})},
\end{aligned}
\end{equation}
where $d_{\rm L}(z)$ denotes the luminosity distance, and the exponential factor accounts for the attenuation of high-energy $\gamma$-rays via electron-positron pair production on the diffuse Extragalactic Background Light (EBL), with the optical depth $\tau$ adopted from Ref.~\cite{Razzaque:2008te}.

Assuming that the intrinsic emission spectra of these sources follow a power-law form characterized by a spectral index $-\Gamma$, the observed differential photon flux is given by (see Appendix \ref{app:details} for a detailed derivation)
\begin{equation}
\frac{\mathrm{d}F}{\mathrm{d}E_{\gamma}} = \frac{L_{\gamma}}{4\pi d_{\rm L}^2} \frac{2-\Gamma}{E_{\gamma,\max}^{2-\Gamma} - E_{\gamma,\min}^{2-\Gamma}} E_{\gamma}^{-\Gamma},
\end{equation}
where we set $E_{\gamma,\min}=0.1~\mathrm{GeV}$ and $E_{\gamma,\max}=1~\mathrm{TeV}$ as the minimum and maximum photon energies reaching the observer from the source, consistent with observations \cite{2003A&A...402..443I, Fermi-LAT:2012nqz, Madejski:2016oqg, Kornecki:2025tej}.

Finally, the effective bias of an unresolved astrophysical source population is formulated as
\begin{equation}
b_{\star}(z) = \int_{L_{\rm min}}^{L_{\rm th}(z)} \mathrm{d}L_{\gamma} \, \phi_{\gamma}(z,L_{\gamma}) \, \frac{b_{\rm h}\left(z,M(L_{\gamma})\right)}{\langle g_{\star}\rangle(z)}.
\end{equation}
Evaluating this integral relies on the connection between the host halo mass $M$ and the intrinsic $\gamma$-ray luminosity $L_{\gamma}$. The specific scaling relations adopted for mAGNs and SFGs are detailed in the subsequent subsections.

\begin{table*}[t]
\setlength{\tabcolsep}{6pt}
\centering
\caption{Best-fit parameters of the IR luminosity function adopted for the three subclasses of star-forming galaxies: spiral galaxies, starburst galaxies, and SF-AGNs. Parameter values are taken from Ref.~\cite{Gruppioni:2013jna}.}
\label{tab:IR_parameters}
\begin{tabular}{lccccccccc}
\toprule
Type & $i$ (index) & $\lambda_{i}$ & $\sigma_{i}$ & $\log_{10}(L^{(i)}_{1}/{\rm L}_{\odot})$ & $\log_{10}(\phi^{(i)}_{1}/\mathrm{Mpc^{-3}})$ & $k^{(i)}_{\rm L}$ & $\kappa_{i}$ & $k^{(i)}_{R_{1}}$ & $k^{(i)}_{R_{2}}$ \\
\midrule
Spiral    & $1$ & 1.0 & 0.50 & 9.78 & $-2.12$ & 4.49 & 1.53 & $-0.54$ & $-7.13$ \\
Starburst & $2$ & 1.0 & 0.35 & 11.17 & $-4.46$ & 1.96 & 2.10 &  3.79 & $-1.06$ \\
SF-AGN    & $3$ & 1.2 & 0.40 & 10.80 & $-3.20$ & 3.17 & 2.10 &  0.67 &  $-3.17$ \\
\bottomrule
\end{tabular}
\end{table*}

\subsubsection{Misaligned AGN}
mAGNs are expected to be one of the dominant contributors to the low-redshift UGRB owing to their large abundance despite their relatively low individual luminosities \cite{DiMauro:2013xta}. However, only a limited number of mAGNs have been detected by current $\gamma$-ray observations, preventing a direct determination of their GLF. Instead, the GLF can be inferred from the much better constrained $5\,{\rm GHz}$ radio core luminosity function through the observed correlation between radio and $\gamma$-ray luminosities \cite{Cholis:2024hmd, Zhou:2024cld}.

Following Refs.~\cite{Cholis:2024hmd, Zhou:2024cld}, the GLF of mAGNs is written as
\begin{equation}\label{GLF_mAGN}
\begin{split}
\phi_{\gamma}(z, L_\gamma) = 
\frac{{\rm d}\ln L_{ \rm 5\,GHz}}
     {{\rm d}\ln L_{\gamma}}
\int & {\rm d}\ln{}L_{ \rm 5\,GHz} \times\\
&
\phi_{ \rm 5\,GHz}(z, L_{ \rm 5\,GHz})
 P(L_{ \rm 5\,GHz}, L_{\gamma}),
\end{split}
\end{equation}
where $\phi_{\rm 5\,GHz}$ denotes the $5\,{\rm GHz}$ radio core luminosity function and $P(L_{\rm 5\,GHz},L_{\gamma})$ describes the intrinsic scatter in the empirical relation between the radio and $\gamma$-ray luminosities. The two luminosities are related through
\begin{equation}
    \log_{10}\left(\frac{L_{\gamma}}{\rm erg\,s^{-1}}\right) = b\log_{10}\left(\frac{L_{\rm 5\,GHz}}{10^{40}\,{\rm erg\,s^{-1}}}\right) + d,
\end{equation}
with parameters $b=0.78$ and $d=40.78$. The radio core luminosity function is parameterized as \cite{2018ApJS..239...33Y}
\begin{equation}\label{eq:RLF_mAGN}
\begin{split}
\phi_{\rm 5\,GHz}(z, L_{\rm 5\,GHz}) & = e_{1}(z)\phi_{1}\\
& \times\Bigg[\left(\frac{L_{\rm 5\,GHz}}{L^{*}\,e_{2}(z)}\right)^{\alpha}
+\left(\frac{L_{\rm 5\,GHz}}{L^{*}\,e_{2}(z)}\right)^{\beta}\Bigg]^{-1},
\end{split}
\end{equation}
with parameters $\log_{10}\left(\phi_{1}/\,{\rm Mpc^{-3}}\right)=-3.749$, $\alpha=0.139$, $\beta=0.878$, and $\log_{10}\left(L^{*}/\,{\rm W\,Hz^{-1}}\right)=21.592$. The redshift evolution is described by
\begin{equation}\label{eq:e1z}
    e_{1}(z) = \frac{(1+z_{\rm c})^{p_{1}}+(1+z_{\rm c})^{p_{2}}}{\left((1+z_{\rm c})/(1+z)
    \right)^{p_{1}}+\left((1+z_{\rm c})/(1+z)\right)^{p_{2}}},
\end{equation}
and
\begin{equation}\label{eq:e2z}
    e_{2}(z) = (1+z)^{k_{1}},
\end{equation}
with $z_{\rm c}=0.893$, $p_{1}=2.085$, $p_{2}=-4.602$, and $k_{1}=1.744$. Meanwhile, the intrinsic scatter entering Eq.~(\ref{GLF_mAGN}) is modeled as 
\begin{eqnarray}
\label{eq:Dispersion_mAGN}
    && P(L_{\rm 5\,GHz}, L_{\gamma}) = \frac{1}{\sqrt{2\pi}\sigma} \nonumber \\
    \times && 
    \exp{\left(-\frac{\left[\log_{10}\left(\frac{L_{\gamma}/{\rm erg\,s^{-1}}}{[L_{\rm 5\,GHz}/{\rm 10^{40}\,erg\,s^{-1}}]^{b}}\right)-d\right]^{2}}{2\sigma^{2}}\right)},
\end{eqnarray}
where $\sigma=0.88$. All parameters entering Eqs.~(\ref{eq:RLF_mAGN})--(\ref{eq:Dispersion_mAGN}) are adopted from Ref.~\cite{2018ApJS..239...33Y} for radio galaxies. 

For the mAGN emission, we assume a power-law photon spectral index of $\Gamma = 2.25$ \cite{Hooper:2016gjy, Blanco:2021icw}. While the intrinsic $\gamma$-ray luminosity of the mAGN population spans the range $10^{40} \le L_{\gamma}/\mathrm{erg\,s^{-1}} \le 10^{50}$, the integration over the GLF to evaluate the unresolved contribution must be truncated at an upper bound $L_{\mathrm{th}}(z)$. The resulting normalized window function for unresolved mAGNs in the $(1.0,\,1.7)\,\mathrm{GeV}$ energy bin is displayed as the dashed green line in the right panel of Fig.~\ref{fig:WindowF_and_PowerS}.

To evaluate the effective bias of mAGN, a relation between halo mass and $\gamma$-ray luminosity is required. Following Ref.~\cite{Arcari:2022zul}, we adopt
\begin{eqnarray}
    M(L_{\gamma}) = \frac{3.6\times10^{13}\,{\rm M_{\odot}}}{(1+z)^{0.903}}\left(\frac{L_{\gamma}}{10^{48}\,{\rm erg\,s^{-1}}}\right)^{0.1032}.
\end{eqnarray}

The cross-power spectrum at $z=0.05$ between galaxies and the $\gamma$-ray flux from unresolved mAGNs within the $(1.0, 1.7)\,\mathrm{GeV}$ energy bin is shown as the dashed green line in the left panel of Fig.~\ref{fig:WindowF_and_PowerS}. On small scales, the cross-power spectrum approaches a nearly scale-independent plateau. This behavior arises because unresolved mAGNs are modeled as point-like emitters with a Dirac delta-function emissivity profile, whose Fourier transform is independent of wavenumber. Consequently, the one-halo contribution is dominated by the central galaxy term, while the scale-dependent satellite contribution remains subdominant, resulting in an approximately constant one-halo power spectrum. This behavior contrasts with the galaxy auto-power spectrum, whose one-halo term is primarily determined by central-satellite and satellite-satellite pairs that trace the halo density profile and therefore retain a pronounced scale dependence \cite{Zheng:2004id, Ando:2014aoa}.

\subsubsection{Star-forming galaxies}
SFGs constitute another important population contributing to the UGRB. Since the $\gamma$-ray emission from SFGs is closely linked to their star-formation activity through cosmic-ray interactions with the interstellar medium, the GLF can be inferred from the infrared (IR) luminosity function, which provides an excellent tracer of the star-formation rate. Following Ref.~\cite{Pinetti:2021jjs}, the GLF is related to the IR luminosity function through
\begin{equation}
    \phi_{\gamma}(z, L_\gamma) = \phi_{\rm IR}\frac{{\rm d}\log_{10}{L_{\rm IR}}}{{\rm d}\log_{10}L_{\gamma}},
\end{equation}
where $L_{\rm IR}$ denotes the infrared luminosity of the galaxy. The empirical relation between the $\gamma$-ray and infrared luminosities is taken to be \cite{Fermi-LAT:2012nqz}
\begin{equation}
    \log_{10}\left(\frac{L_{\gamma}}{\rm erg\,s^{-1}}\right) = \alpha_{\rm IR}\log_{10}\left(\frac{L_{\rm IR}}{10^{10}\,{\rm L}_{\odot}}\right) + \beta_{\rm IR},
\end{equation}
with $\alpha_{\rm IR} = 1.09$ and $\beta_{\rm IR} = 39.19$.

To model the IR luminosity function, we adopt the framework developed by Gruppioni et al.~\cite{Gruppioni:2013jna}, in which the SFG population is decomposed into three subclasses, quiescent spiral galaxies, starburst galaxies, and star-forming galaxies hosting obscured or low-luminosity active galactic nuclei (SF-AGN). The total IR luminosity function is therefore expressed as
\begin{equation}
\phi_{\rm IR} = \sum^{3}_{i=1}\phi_{i},
\end{equation}
where each component (Table~\ref{tab:IR_parameters} for $i=1, 2, 3$) is parameterized as \cite{Gruppioni:2013jna, Pinetti:2021jjs}
\begin{equation}
    \phi_{i} = \phi^{(i)}_{0}(z)\left(\frac{L_{\rm IR}}{L^{(i)}_{0}}\right)^{1-\lambda_{i}}\exp\left[-\frac{1}{2\sigma_{i}^{2}}\log_{10}^{2}\left(1+\frac{L_{\rm IR}}{L^{(i)}_{0}}\right)\right],
\end{equation}
where $\phi^{(i)}_{0}$ is defined piecewise
\begin{equation}
    \phi^{(i)}_{0} = 
    \begin{cases}
        \phi^{(i)}_{1}\left(({1+z})/{1.15}\right)^{k^{(i)}_{R_{1}}} & {\rm for} \, z\leq0.53, \\
\phi^{(i)}_{1}\left({\kappa_{i}}/{1.15}\right)^{k^{(i)}_{R_{1}}}\left[({1+z})/{\kappa_{i}}\right]^{k^{(i)}_{R_{2}}} & {\rm for}\,z>0.53.
    \end{cases}
\end{equation}
Similarly, $L^{(i)}_{0}$ is given by
\begin{equation}
L^{(i)}_{0} = 
\begin{cases}
L^{(i)}_{1}\left(({1+z})/{1.15}\right)^{k^{(i)}_{L}} & {\rm for}\, z\leq1.1, \\
L^{(i)}_{1}\left({2.1}/{1.15}\right)^{k^{(i)}_{L}} & {\rm for}\, z > 1.1.
\end{cases}
\end{equation}
The corresponding parameter values are summarized in Table~\ref{tab:IR_parameters} following Ref.~\cite{Gruppioni:2013jna}. For the SFG population, we adopt a photon spectral index of $\Gamma = 2.7$ \cite{Pinetti:2019ztr} and perform the integration over the luminosity range $10^{37} \le L_{\gamma}/\mathrm{erg\,s^{-1}} \le \min[10^{42}, L_{\mathrm{th}}(z)]$. The resulting normalized window function for unresolved SFGs within the $(1.0,\,1.7)\,\mathrm{GeV}$ energy bin is depicted as the dash-dotted brown curve in the right panel of Fig.~\ref{fig:WindowF_and_PowerS}. Crucially, its characteristic multi-peaked profile imprints the distinct cosmic evolution profiles of the three underlying SFG subclasses, quiescent spiral galaxies predominantly drive the signal at low redshifts ($z \lesssim 0.5$), SF-AGN emerge as the dominant population at intermediate cosmic epochs with a broad maximum anchoring around $z \simeq 1.1$, whereas starburst galaxies yield a comparatively subdominant contribution across the entire redshift range under consideration.

The mass-to-luminosity relation for SFG is \cite{Arcari:2022zul}
\begin{equation}
M(L_{\gamma}) = \frac{10^{12}\,\mathrm{M}_{\odot}}{(1+z)^{1.61}} \left( \frac{L_{\gamma}}{6.8 \times 10^{39}\,\mathrm{erg\,s^{-1}}} \right)^{0.92}.
\end{equation}

The corresponding cross-power spectrum evaluated at $z = 0.05$ between the galaxy and the $\gamma$-ray flux from unresolved SFGs in the $(1.0, 1.7)\,\mathrm{GeV}$ energy bin is displayed as the dash-dotted brown line in the left panel of Fig.~\ref{fig:WindowF_and_PowerS}.

\subsubsection{ALP-CR proton-induced $\gamma$-ray emission}
In this work, we consider $\gamma$-ray production through the scattering between cold ALP dark matter and relativistic CR protons
\begin{equation}
a + p \rightarrow p + \gamma,
\end{equation}
following the framework developed in Ref.~\cite{Goncalves:2026ean}. Throughout this work, we focus on photophobic ALPs, for which couplings to photons are assumed to be strongly suppressed. Consequently, the interaction is taken to proceed exclusively through the ALP-proton coupling. The corresponding effective interaction is described by the Lagrangian \cite{Lella:2024dmx}
\begin{equation}
\mathcal{L} = \frac{1}{2} \frac{g_{a\mathrm{p}}}{m_{\mathrm{p}}} (\partial_\mu a) \bar{p} \gamma^\mu \gamma_5 p,
\end{equation}
where $g_{a\mathrm{p}}$ denotes the ALP-proton coupling parameter, $m_{\mathrm{p}}$ is the proton mass, and $a$ and $p$ represent the ALP and proton fields, respectively.

To evaluate the $\gamma$-ray emission generated by ALP-CR proton interactions across the cosmological halo population, a description of the CR proton population within halos over cosmological volumes is required. Since the CR distribution cannot be directly probed for the entire halo population, we construct a phenomenological bottom-up model that connects the steady-state CR population to the global properties of each host dark matter halo. The underlying physical premise is that GeV-PeV CRs are predominantly accelerated by supernova remnants \cite{Gabici:2019jvz, Hanasz:2021rfs}. Consequently, the CR injection power is ultimately governed by the star-formation activity of the host galaxy \cite{Fermi-LAT:2010zba}.

To enable the CR normalization to be determined self-consistently from the halo mass $M$, our model establishes a multi-step phenomenological mapping
\[
M \longrightarrow M_* \longrightarrow \mathrm{SFR} \longrightarrow \dot{N}_{\mathrm{SN}} \longrightarrow L_{\mathrm{p}},
\]
where $M_{*}$, $\mathrm{SFR}$, $\dot{N}_{\mathrm{SN}}$, and $L_{\mathrm{p}}$ denote the total stellar mass, the star-formation rate, the core-collapse supernova rate, and the CR proton injection luminosity, respectively.

To connect the stellar component to the host dark matter halo, we adopt the stellar-to-halo mass relation (SHMR) of Ref.~\cite{Girelli:2020goz}
\begin{equation}
    \frac{M_{*}}{M}(z) = 2B(z)\left[\left(\frac{M}{M_{B}(z)}\right)^{-\beta(z)} + \left(\frac{M}{M_{B}(z)}\right)^{\gamma(z)}\right]^{-1},
\end{equation}
where the redshift-dependent quantities are defined as
\begin{equation}
\begin{aligned}
    & B(z) = C(1+z)^{\phi} ,\\
    & \log_{10}M_{B}(z) = D +z\cdot\mu , \\
    & \beta(z) = E\cdot z + F, \\
    &\gamma(z) = G(1+z)^{\eta}.
\end{aligned}
\end{equation}
The specific parameter values adopted in this work are summarized in Table~\ref{SHMR_table}.

The SFR is related to the total stellar mass through the star-forming main sequence \cite{2014ApJS..214...15S}
\begin{equation}
\begin{aligned}
    \log_{10} \left( {\rm SFR}\left[\rm M_{\odot}\,yr^{-1}\right]\right) & = (0.84 - 0.026t)\log_{10}\left(M_{*} / {\rm M_{\odot}}\right) \\
    &\quad- (6.51 - 0.11t),
\end{aligned}
\end{equation}
where $t$ denotes the cosmic age in units of $\mathrm{Gyr}$.

Assuming a standard Salpeter initial mass function (IMF), the corresponding core-collapse supernova rate is
\begin{equation}
\dot{N}_{\mathrm{SN}} \left[\mathrm{yr}^{-1}\right] = \frac{\int_{8\,\mathrm{M}_{\odot}}^{100\,\mathrm{M}_{\odot}} \phi(M_{\rm star})\,\mathrm{d}M_{\rm star}}{\int_{0.1\,\mathrm{M}_{\odot}}^{100\,\mathrm{M}_{\odot}} M_{\rm star} \phi(M_{\rm star})\,\mathrm{d}M_{\rm star}} \times \mathrm{SFR}.
\end{equation}
For a Milky Way-like halo ($M \sim 10^{12}\,\mathrm{M}_\odot$), our model predicts a supernova rate of approximately $0.01\,\mathrm{yr}^{-1}$, which is in excellent agreement with Galactic observations \cite{Diehl:2006cf}.

\begin{table}[t]
    \centering
    \caption{Best-fit parameters of the SHMR used in this work. The parameter values are adopted from Ref.~\cite{Girelli:2020goz}.}
    \label{SHMR_table}
    \addtolength{\tabcolsep}{1pt} 
    \begin{tabular}{l ccc cccc c}
        \toprule 
         & $C$ & $\phi$ & $D$ & $\mu$ & $E$ & $F$ & $G$ & $\eta$ \\
        \midrule
        Best fit & 0.046 & $-0.38$ & 11.79 & 0.20 & 0.043 & 0.96 & 0.709 & $-0.18$ \\
        \bottomrule
    \end{tabular}
\end{table}

Each supernova is assumed to release a mechanical energy of $E_{\mathrm{SN}} = 10^{51}\,\mathrm{erg}$, of which a fraction $\eta_{\mathrm{CR}} = 0.1$ is converted into cosmic rays \cite{Strong:2010pr}. Empirically, CR protons compose the overwhelming majority ($\approx 98.6\%$) of the total hadronic and electronic CR energy budget at injection \cite{Strong:2010pr}. We therefore adopt the well-justified approximation that the accelerated CR energy is channeled entirely into the proton component. Neglecting radiative losses for these relativistic protons, whose cooling times are much longer than their escape times in the halos considered here, the steady-state CR proton injection luminosity simplifies to
\begin{equation}
L_{\mathrm{p}} = \eta_{\mathrm{CR}} E_{\mathrm{SN}} \dot{N}_{\mathrm{SN}}.
\end{equation}

Within our theoretical framework, we assume that the propagated CR proton spectrum shares a universal spectral shape among star-forming halos, while the overall normalization $A_{\mathrm{p}}$ is strictly determined by the halo-specific injection power. The propagated steady-state CR proton differential flux is written as
\begin{equation}
\frac{\mathrm{d}\Phi_{\mathrm{p}}}{\mathrm{d}E_{\mathrm{p}}} = A_{\mathrm{p}} f_{\mathrm{p}}(E_{\mathrm{p}}) f_{\mathrm{p}}(\mathbf{x}),
\end{equation}
where $A_{\mathrm{p}}$ has units of $\mathrm{cm}^{-2}\,\mathrm{s}^{-1}\,\mathrm{sr}^{-1}\,\mathrm{GeV}^{-1}$. The spectral profile $f_{\mathrm{p}}(E_{\mathrm{p}})$ for GeV-PeV protons is modeled as \cite{Olinto:2004kn, Aharonian:2026tzf}
\begin{equation}
f_{\mathrm{p}}(E_{\mathrm{p}}) = \left( \frac{E_{\mathrm{p}}}{E_{\mathrm{p, scale}}} \right)^{-2.7},
\end{equation}
with the pivot energy $E_{\mathrm{p, scale}} = 1\,\mathrm{TeV}$. Moderate variations in this spectral index are found to have a negligible impact on our final forecasts. 

The spatial distribution of cosmic rays within the halo is modeled based on the numerical simulations of Ref.~\cite{Ramesh:2024bvf}. By performing a phenomenological fit to their simulated distributions, we find that the steady-state CR spatial profile can be approximately parameterized by
\begin{equation}
f_{\mathrm{p}}(\mathbf{x}) = \frac{1}{\left(\frac{r}{R_{\mathrm{200c}}}\right)\left(1 + \frac{r}{R_{\mathrm{200c}}}\right)^{2}},
\end{equation}
where $R_{\mathrm{200c}}(z, M)$ represents the halo radius enclosing an average density equal to 200 times the critical density of the Universe.

The normalization $A_{\mathrm{p}}$ is uniquely fixed by requiring the total steady-state CR proton energy confined within the halo to balance the accumulated injection over the effective escape time
\begin{equation}
E_{\mathrm{CR,p}} = L_{\mathrm{p}} \times \tau_{\mathrm{esc}}.
\end{equation}
We approximate $\tau_{\mathrm{esc}}$ using the diffusion timescale \cite{Hanasz:2021rfs, Pandey:2024cib}
\begin{equation}
\tau_{\mathrm{esc}}(z, M) \simeq \frac{R_{\mathrm{vir}}^{2}(z, M)}{D_{0}},
\end{equation}
where we assume a constant diffusion coefficient and adopt a characteristic Milky Way-like value,
$D_{0}\sim10^{28}\,\mathrm{cm^{2}\,s^{-1}}$, as the fiducial diffusion coefficient. We will address the uncertainties associated with $D_0$ in next section. The normalization $A_{\mathrm{p}}(z, M)$ is thus computed by integrating over the halo volume and CR proton energy
\begin{equation}
L_{\mathrm{p}} \times \tau_{\mathrm{esc}} = \int_{\rm 1\, GeV}^{\rm 10\, PeV} \mathrm{d}E_{\mathrm{p}} E_{\mathrm{p}} \int^{R_{\rm vir}} \mathrm{d}^{3}\mathbf{x} \frac{4\pi}{v_{\mathrm{p}}} \frac{\mathrm{d}\Phi_{\mathrm{p}}}{\mathrm{d}E_{\mathrm{p}}},
\end{equation}
where $v_{\mathrm{p}} \approx c$ is the CR proton velocity.

The window function $W_{\mathrm{ap}}$, characterizing the contribution from the ALP-CR proton interaction to the observed $\gamma$-ray intensity at energy $E_{\gamma}$ from redshift $z$, is derived as (see Appendix \ref{App:Axion_WF} for the full derivation)
\begin{equation}\label{eq:WF_Axion}
\begin{aligned}
    W_{\mathrm{ap}}(E_{\gamma},\chi) = & \frac{\langle g_{a}\rangle(z)}{m_{a}(1+z)^{3}} \int_{E^{\mathrm{min}}_{\mathrm{p}}([1+z]E_{\gamma})} ^{10\,{\rm PeV}}\mathrm{d}E_{\mathrm{p}} f_{\mathrm{p}}(E_{\mathrm{p}}) \times \\
    &\frac{\mathrm{d}\sigma_{a+\mathrm{p} \rightarrow \mathrm{p} + \gamma}}{\mathrm{d}E_{\gamma}}([1+z]E_{\gamma}, E_{\mathrm{p}}, m_{a}) e^{-\tau([1+z]E_{\gamma})},
\end{aligned}
\end{equation}
where $m_{a}$ denotes ALP DM mass.

The spatial source field is defined as
\begin{equation}
g_{a}(\mathbf{x}|z, M) = \rho_{\rm DM}(\mathbf{x}|z, M) A_{\mathrm{p}}(z, M) f_{\mathrm{p}}(\mathbf{x}),
\end{equation}
with its ensemble average given by
\begin{equation}
\langle g_{a}\rangle(z) = \int \mathrm{d}M \frac{\mathrm{d}n}{\mathrm{d}M} A_{\mathrm{p}}(z,M) \int \mathrm{d}^{3}\mathbf{x}\,\rho_{\rm DM}(\mathbf{x}|z, M) f_{\mathrm{p}}(\mathbf{x}).
\end{equation}
The kinematic threshold $E_{\mathrm{p}}^{\mathrm{min}}(E_{\gamma})$, which is the minimum proton energy required to produce a $\gamma$-ray of energy $E_{\gamma}$ in the rest frame, is given by \cite{Dent:2020qev}
\begin{equation}
E_{\mathrm{p}}^{\mathrm{min}}(E_{\gamma}) = \frac{E_{\gamma} - m_{a}}{2} + \frac{E_{\gamma}}{2}\sqrt{1 + \frac{4m_{\mathrm{p}}^{2}}{m_{a}(2E_{\gamma} - m_{a})}}.
\end{equation}

The normalized window function for the $(1.0, 1.7)\,\mathrm{GeV}$ energy bin, induced by an ALP-CR proton interaction with mass $m_{a} = 1\,\mathrm{MeV}$, is shown as the solid black curve in the right panel of Fig.~\ref{fig:WindowF_and_PowerS}.

The Fourier transform of the field $g_{a}$ is then given by 
\begin{equation}
\begin{aligned}
    \tilde{g}_{a}(k|z, M) & = \int {\rm d}^{3}\mathbf{x}\,g_{a}(\mathbf{x}|z, M)e^{-i\mathbf{k}\cdot \mathbf{x}}\\
    & = \begin{aligned}[t]
        & 4\pi\rho_{\rm s}(z, M)A_{\rm p}(z,M)r_{\rm s}^{3}(z,M)R_{\rm 200c}^{3}(z, M) \\
        &\times \int_{0}^{R_{\rm vir}}\frac{{\rm d}r}{\left(r_{\rm s} + {r}\right)^{2}\left(R_{\rm 200c} + {r}\right)^{2}}\frac{\sin{kr}}{kr}.
    \end{aligned}
\end{aligned}
\end{equation}
The scale-independent bias is then defined as
\begin{equation}
    b_{\gamma}(z) = \int{\rm d}M \frac{{\rm d}n}{{\rm d}M}b_{\rm h}(z, M)\int{\rm d}^{3}\mathbf{x}\frac{g_{a}(\mathbf{x}|z, M)}{\langle g_{a} \rangle(z)}.
\end{equation}

The resulting cross-power spectrum of the ALP-CR proton-induced $\gamma$-ray signal with the galaxy catalog at $z=0.05$ is shown as the solid black curve in the left panel of Fig.~\ref{fig:WindowF_and_PowerS}.

\begin{figure}
    \centering
    \includegraphics[width=1.\linewidth]{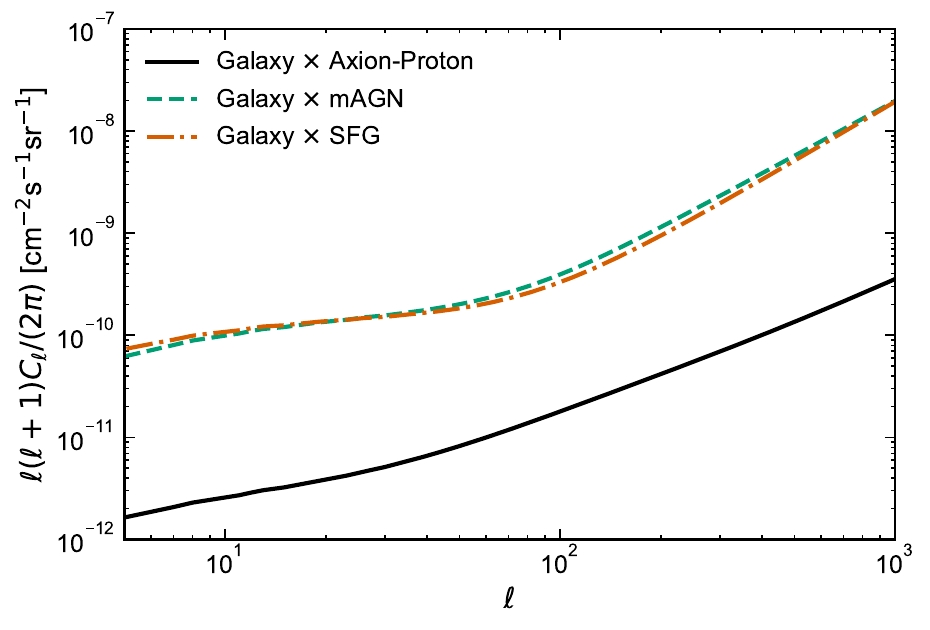}
    \caption{Predicted cross-angular power spectra between the 2MRS galaxy catalog and $\gamma$-ray emission from ALP-CR proton interactions (solid black; $m_{a}=1\,{\rm MeV}$ and $g_{\rm ap}=1$), unresolved mAGNs (dashed green), and unresolved SFGs (dash-dotted brown) in the $(1.0, 1.7)\,{\rm GeV}$ energy bin.}
    \label{fig:APS}
\end{figure}

\section{Sensitivity Forecast and Discussion}\label{Sec:ResultDiscussion}
Utilizing the window functions and three-dimensional power spectra derived in the preceding section, we calculate the cross APS between the sources contributing to the UGRB and the 2MRS galaxy catalog via Eq.~(\ref{eq:APS}).

Figure~\ref{fig:APS} displays the predicted cross APS integrated over the redshift interval $0 \le z\le 0.1$ within the $(1.0, 1.7)\,\mathrm{GeV}$ energy bin. The contributions from unresolved mAGNs (dashed green), SFGs (dash-dotted brown), and the ALP-CR proton-induced $\gamma$-ray signal (solid black), assuming a benchmark ALP mass of $m_a=1\,\mathrm{MeV}$ and coupling $g_{\rm ap}=1$, are presented separately. As expected, conventional astrophysical sources strictly dominate the observable anisotropy across the entire multipole range, whereas the ALP-CR proton-induced component remains subdominant.

To estimate the statistical uncertainty of the predicted cross APS, we assume Gaussian statistics, under which the covariance is given by
\begin{equation}
    \left(\Delta C_{\ell}^{\gamma\mathrm{g}}\right)^{2} = \frac{f_{\mathrm{sky}}^{-1}}{2\ell + 1} \left[ \left(C_{\ell}^{\gamma\mathrm{g}}\right)^{2} + \left(C_{\ell}^{\gamma\gamma} + \frac{C_{\mathrm{N}}^{\gamma}}{\left(B_{\ell}^{\gamma}\right)^{2}}\right) \left(C_{\ell}^{\mathrm{gg}} + C_{\mathrm{N}}^{\mathrm{g}}\right) \right],
\end{equation}
where $f_{\mathrm{sky}} = \min\{f_{\mathrm{2MRS}}, f_{\mathrm{Fermi}}\}$ denotes the effective overlapping sky coverage fraction. Here, $C_{\ell}^{\gamma\gamma}$ and $C_{\ell}^{\mathrm{gg}}$ represent the auto-correlation APS of the $\gamma$-ray sky and the galaxy distribution, respectively. The term $C_{\mathrm{N}}^{\gamma}$ accounts for the photon Poisson noise (summarized in Table~\ref{tab:FermiLAT_specifications}), and $C_{\mathrm{N}}^{\mathrm{g}} = 4\pi f_{\mathrm{2MRS}} / N_{\mathrm{g}}$ specifies the galaxy shot noise.

The finite angular resolution of the \textit{Fermi}-LAT instrument suppresses the observed anisotropy on small angular scales. We incorporate this observational effect using the energy-dependent beam window function
\begin{equation}
    B_{\ell}^{\gamma}(E_\gamma) = \exp \left[ -\frac{\theta_b^2(E_\gamma, \ell) \ell^2}{2} \right],
\end{equation}
where the effective beam size $\theta_b(E_\gamma, \ell)$ is parameterized by
\begin{equation}
    \theta_b(E_\gamma, \ell) = \theta_{\mathrm{cont}}^{68\%}(E_\gamma) \left[ 1 + 0.25 \, \theta_{\mathrm{cont}}^{68\%}(E_\gamma) \ell \right]^{-1}.
\end{equation}
Following Refs.~\cite{Pinetti:2019ztr,Zhou:2024cld}, the $68\%$ containment angle is given by
\begin{equation}
    \theta_{\mathrm{cont}}^{68\%}(E_\gamma) = \theta_{\mathrm{ref}}^{68\%} \left( \frac{E_\gamma}{E_{\mathrm{ref}}} \right)^{-0.95} + 0.05^\circ ,
\end{equation}
with parameters $E_{\mathrm{ref}} = 0.5\,\mathrm{GeV}$ and $\theta_{\mathrm{ref}}^{68\%} = 1.20\,{\rm deg}$. Given that the angular resolution of galaxy surveys is substantially superior to that of $\gamma$-ray telescopes, no beam correction is required for the galaxy density field.

To evaluate the projected sensitivity to the ALP-proton coupling, we perform a statistical forecast. The null hypothesis assumes that the observable signal consists entirely of unresolved astrophysical $\gamma$-ray background, whereas the alternative hypothesis additionally incorporates the ALP-CR proton-induced $\gamma$-ray emission.

Following the approach in Ref.~\cite{Pinetti:2019ztr, Zhou:2024cld}, for a given ALP mass, the statistical significance is quantified by the test statistic $\Delta\chi^2$, defined as
\begin{equation}
\Delta\chi^2
=
\sum_{\ell,e}
\left(
\frac{C_{\ell,e}^{\gamma g}}{\Delta C_{\ell,e}^{\gamma g}}
\right)^2_{\rm \star+ALP}
-
\sum_{\ell,e}
\left(
\frac{C_{\ell,e}^{\gamma g}}{\Delta C_{\ell,e}^{\gamma g}}
\right)^2_{\star},
\end{equation}
where the summation runs over all angular multipoles $\ell$ and $\gamma$-ray energy bins $e$. 

\begin{figure}
    \centering
    \includegraphics[width=1.\linewidth]{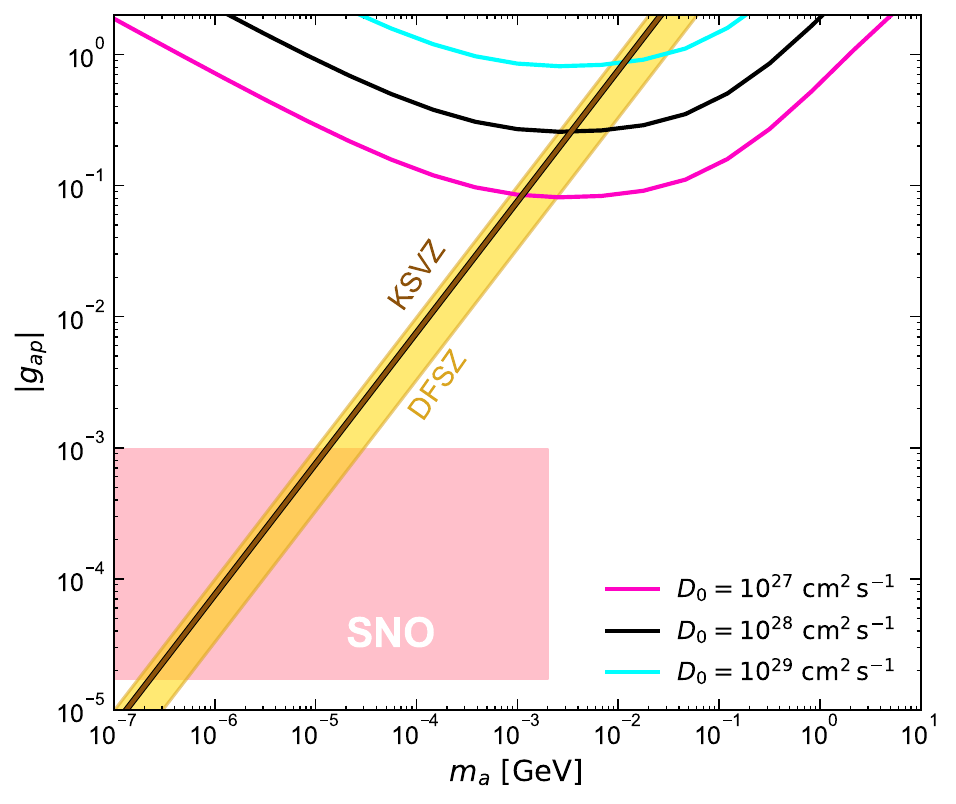}
    \caption{Forecasted $2\sigma$ ($95.4\%$) sensitivity to the ALP-proton coupling $g_{\rm ap}$ as a function of the ALP DM mass $m_{a}$. Curves correspond to different assumptions for the cosmic-ray diffusion coefficient $D_{0}$, with the solid black curve ($D_{0} = 10^{28}\,{\rm cm^{2}s^{-1}}$) representing the fiducial model. The projected sensitivity is compared with the existing constraint from the Sudbury Neutrino Observatory (SNO) \cite{Bhusal:2020bvx}. The yellow band indicates the parameter space predicted by representative QCD axion models.}
    \label{fig:UpperLimits}
\end{figure}

Since the ALP-proton coupling $g_{\rm ap}$ is the sole additional free parameter for a fixed ALP mass, the test statistic $\Delta\chi^2$ asymptotically follows a $\chi^2$ distribution with one degree of freedom. Consequently, we establish $\Delta\chi^2 = 4$ to derive the projected $2\sigma$ ($95.4\%$ confidence level) exclusion limits. The resulting projected constraints on $g_{\rm ap}$ are presented in Fig.~\ref{fig:UpperLimits}, alongside existing experimental bounds from the Sudbury Neutrino Observatory (SNO) \cite{Bhusal:2020bvx}.

The primary theoretical uncertainty in our sensitivity forecast stems from the modeling of CR propagation within dark matter halos. Specifically, the steady-state CR proton abundance scales with their residence time, which is inversely governed by the diffusion coefficient $D_0$. Both observational constraints and numerical simulations indicate that $D_0$ can vary by several orders of magnitude across different galactic environments \cite{Strong:2007nh, Wiener:2016zcr, Dorner:2022rap, Pandey:2024cib, Ramesh:2024bvf}. To assess the impact of this uncertainty, we repeat the forecast by varying the diffusion coefficient within the range $D_0 = 10^{27}\text{--}10^{29}\,\mathrm{cm^2\,s^{-1}}$. As anticipated, a larger diffusion coefficient accelerates CR escape, thereby diminishing the steady-state CR density, reducing the ALP-CR proton-induced $\gamma$-ray emissivity, and leading to milder constraint bounds. Nevertheless, even under conservative propagation assumptions (i.e., maximal diffusion coefficient), our cross-correlation technique retains the capability to probe previously unconstrained regions of the ALP parameter space.

We emphasize that the CR model adopted here serves as a first-order phenomenological representation of the ensemble-averaged properties of star-forming halos. While individual galaxies exhibit local variations in CR distributions due to distinct magnetic field topologies, radiative feedback, and star-formation histories, these micro-level fluctuations are expected to average out over the cosmological halo population considered in this work. A more detailed modeling framework utilizing cosmological magnetohydrodynamic (MHD) simulations with explicit CR transport will constitute an intriguing direction for future work.

Furthermore, the ALP-CR proton-induced $\gamma$-ray window function shown in the right panel of Fig.~\ref{fig:WindowF_and_PowerS} exhibits a prominent peak at intermediate redshifts ($z \sim 1$). Consequently, future cross-correlation analyses leveraging next-generation galaxy catalogs with deeper redshift coverage (e.g., DESI, Euclid, or LSST) will significantly boost the detection sensitivity to ALP-CR proton-induced signatures. At these higher redshifts, however, additional unresolved source populations, most notably BL Lac objects and Flat Spectrum Radio Quasars, will become increasingly non-negligible and must be integrated into the background covariance.

Finally, we highlight a fundamental distinction between the ALP-CR proton-induced $\gamma$-ray signal studied here and conventional dark matter signatures. Standard dark matter decay or annihilation signals (especially for WIMP dark matter \cite{Ando:2005xg, Fornengo:2013rga}) depend solely on the dark matter density field. In contrast, the signal investigated in this work arises from the interplay between two distinct physical fields, namely the dark matter profile and the ambient CR proton distribution. Consequently, the effective spatial source distribution entering the halo model is modified, rendering the predicted spatial clustering sensitive to the joint distribution of both dark matter and cosmic rays within halos. This offers a novel theoretical link between particle dark matter models and large-scale structure observations, establishing a complementary avenue for probing non-gravitational dark matter interactions on cosmological scales.

\section{Conclusion}\label{Sec:Conclusion}

In this work, we have investigated the cosmological $\gamma$-ray signal produced by interactions between axion-like particle dark matter and $\mathrm{GeV}$--$\mathrm{PeV}$ cosmic-ray protons, and explored the capability of large-scale structure cross-correlations to probe the ALP-proton coupling. Assuming that ALPs constitute the dark matter component of the Universe, we modeled the diffuse $\gamma$-ray emission generated through the scattering process $a+p\rightarrow p+\gamma$ inside dark matter halos. The resulting contribution to the unresolved $\gamma$-ray background was incorporated into a halo-model framework to forecast the sensitivity of the cross-correlation between the \textit{Fermi}-LAT UGRB and the 2MRS galaxy catalog.

Unlike conventional indirect dark matter searches, where the source field is determined solely by the dark matter distribution, the observable considered here originates from the interaction between two distinct cosmic fields, namely the dark matter density and the CR proton density. Consequently, the effective source field is proportional to the product of these two components, leading to qualitatively different clustering behavior and establishes a new framework for connecting particle dark matter interactions with large-scale structure observables.

To model the cosmological CR population, we adopted a phenomenological bottom-up approach in which the CR injection rate is linked to the star formation rate through the supernova rate, while the steady-state CR density is determined by diffusive escape. Within this framework, the ALP-CR proton-induced $\gamma$-ray emission from the cosmological halo population was computed self-consistently and incorporated into the halo-model calculation of the UGRB-galaxy cross-correlation. Our forecasts show that this approach is capable of probing previously unexplored regions of the ALP parameter space over a broad ALP mass range.

The dominant theoretical uncertainty arises from the modeling of CR transport within dark matter halos, particularly the diffusion coefficient and the spatial distribution of CRs, both of which are expected to vary among different galactic environments. The phenomenological model adopted here should therefore be regarded as describing the ensemble-averaged properties of the halo population rather than individual galaxies. Future cosmological simulations incorporating CR transport and magnetic fields will provide important guidance for reducing these uncertainties.

More generally, the formalism developed in this work can be readily extended to other dark matter scenarios in which observable signals arise from interactions between dark matter and additional astrophysical fields. Future large-scale structure surveys, together with deeper $\gamma$-ray observations and multi-tracer cross-correlation analyses, will further enhance the sensitivity of this approach and provide new opportunities to explore particle dark matter using cosmological observables.

\section{Acknowledgments}
The financial assistance of the South African Radio Astronomy Observatory (SARAO) towards this research is hereby acknowledged (\url{www.sarao.ac.za}).

\clearpage

\appendix
\onecolumngrid

\section{Calculation Details for Astrophysical Sources} \label{app:details}
In the observer frame, the differential $\gamma$-ray flux from a source is conventionally parameterized as a function of the observed photon energy $E_\gamma$, following a power-law form
\begin{equation}
    \frac{\mathrm{d}F}{\mathrm{d}E_\gamma} = N_{0}\left(\frac{E_\gamma}{E_{\gamma, \min}}\right)^{-\Gamma},
\end{equation}
where $N_{0}$ represents the normalization factor, $\Gamma$ is the spectral index determined by the underlying emission mechanism, and $E_{\gamma, \min}$ denotes the lower bound of the photon energy reaching the observer.

The intrinsic $\gamma$-ray energy luminosity of the source, defined in its rest frame, is given by the integration of the differential rest-frame luminosity
\begin{equation}\label{L_rest}
    L_{\gamma} = \int_{E_{\gamma, \min}^{\mathrm{r}}}^{E_{\gamma, \max}^{\mathrm{r}}} \mathrm{d}E^{\mathrm{r}}_\gamma \frac{\mathrm{d}L_{\gamma}}{\mathrm{d}E^{\mathrm{r}}_\gamma},
\end{equation}
where $E^{\mathrm{r}}_\gamma = (1+z)E_\gamma$ accounts for the cosmological redshift $z$, and $E^{\mathrm{r}}_{\gamma, \min/\max}$ correspond to the boundary energies emitted in the source rest frame. The rest-frame differential energy luminosity is related to the observer-frame differential photon flux via the standard cosmological distance relation
\begin{equation}
    \frac{\mathrm{d}L_{\gamma}}{\mathrm{d}E^{\mathrm{r}}_\gamma} = \frac{4\pi d_{\mathrm{L}}^{2}(z)}{1+z} E_\gamma \frac{\mathrm{d}F}{\mathrm{d}E_\gamma},
\end{equation}
where $d_{\mathrm{L}}(z)$ is the luminosity distance.

Substituting this relation into Eq.~(\ref{L_rest}) and transforming the integration variable to the observer frame, we obtain
\begin{equation}
\begin{aligned}
    L_{\gamma} &= \int_{E_{\gamma, \min}^{\mathrm{r}}}^{E_{\gamma, \max}^{\mathrm{r}}} \mathrm{d}E^{\mathrm{r}}_{\gamma} \frac{4\pi d_{\mathrm{L}}^{2}(z)}{1+z} E_\gamma \frac{\mathrm{d}F}{\mathrm{d}E_\gamma} \\
    &= 4\pi d_{\mathrm{L}}^{2}(z) \int_{E_{\gamma, \min}}^{E_{\gamma, \max}} \mathrm{d}E_\gamma \, E_\gamma \frac{\mathrm{d}F}{\mathrm{d}E_\gamma} \\
    &= 4\pi d_{\mathrm{L}}^{2}(z) N_{0} \int_{E_{\gamma, \min}}^{E_{\gamma, \max}} \mathrm{d}E_\gamma \, E_\gamma \left(\frac{E_\gamma}{E_{\gamma, \min}}\right)^{-\Gamma} \\
    &= 4\pi d_{\mathrm{L}}^{2}(z) N_{0} E_{\gamma, \min}^{\Gamma} \frac{E_{\gamma, \max}^{2-\Gamma} - E_{\gamma, \min}^{2-\Gamma}}{2-\Gamma}.
\end{aligned}
\end{equation}
Consequently, the normalization factor $N_0$ can be self-consistently solved as
\begin{equation}\label{N0}
    N_{0} = \frac{L_{\gamma}}{4\pi d_{\mathrm{L}}^{2}(z)} \frac{2-\Gamma}{E_{\gamma, \min}^{\Gamma}} \frac{1}{E_{\gamma, \max}^{2-\Gamma} - E_{\gamma, \min}^{2-\Gamma}}.
\end{equation}

By inserting Eq.~(\ref{N0}) back into the initial power-law spectrum, the differential photon flux simplifies to
\begin{equation}
    \frac{\mathrm{d}F}{\mathrm{d}E_\gamma} = \frac{L_{\gamma}}{4\pi d_{\mathrm{L}}^{2}(z)} \frac{2-\Gamma}{E_{\gamma, \max}^{2-\Gamma} - E_{\gamma, \min}^{2-\Gamma}} E_{\gamma}^{-\Gamma}.
\end{equation}
Here again, $E_{\gamma, \min}$ and $E_{\gamma, \max}$ denote the minimum and maximum photon energies reaching the observer from the source.

When observed by a telescope, the integrated photon flux is attenuated by the extragalactic background light, characterized by the optical depth $\tau$. The total detected photon flux within the telescope's operational energy window $(E_{\mathrm{obs}}^{\min}, E_{\mathrm{obs}}^{\max})$ is expressed as
\begin{equation}
    F_{\mathrm{obs}} = \int_{E_{\mathrm{obs}}^{\min}}^{E_{\mathrm{obs}}^{\max}} \mathrm{d}E_{\gamma} \frac{\mathrm{d}F}{\mathrm{d}E_{\gamma}} e^{-\tau[(1+z)E_{\gamma}, z]}.
\end{equation}

A source is considered detectable if its integrated photon flux reaches the telescope's flux sensitivity threshold, i.e., $F_{\mathrm{obs}} = F_{\mathrm{sens}}$. Substituting the parameterized differential flux into the threshold condition yields
\begin{equation}
    F_{\mathrm{sens}} = \frac{L_{\mathrm{th}}}{4\pi d_{\mathrm{L}}^{2}(z)} \frac{2-\Gamma}{E_{\gamma, \max}^{2-\Gamma} - E_{\gamma, \min}^{2-\Gamma}} \int_{E_{\mathrm{obs}}^{\min}}^{E_{\mathrm{obs}}^{\max}} \mathrm{d}E_{\gamma} E_{\gamma}^{-\Gamma} e^{-\tau[(1+z)E_{\gamma}, z]}.
\end{equation}
Finally, by isolating the threshold luminosity, we arrive at the operational definition of the minimum detectable $\gamma$-ray luminosity $L_{\mathrm{th}}$ for a given instrument
\begin{equation}
    L_{\mathrm{th}} = 4\pi d_{\mathrm{L}}^{2}(z) F_{\mathrm{sens}} \left( \frac{E_{\gamma, \max}^{2-\Gamma} - E_{\gamma, \min}^{2-\Gamma}}{2-\Gamma} \right) \left[ \int_{E_{\mathrm{obs}}^{\min}}^{E_{\mathrm{obs}}^{\max}} \mathrm{d}E_{\gamma} E_{\gamma}^{-\Gamma} e^{-\tau[(1+z)E_{\gamma}, z]} \right]^{-1},
\end{equation}
where we take $F_{\mathrm{sens}}=10^{-10}\,{\rm cm^{-2}s^{-1}}$ as the photon flux sensitivity of the Fermi-LAT \cite{Pinetti:2019ztr}.

\section{Window Function for ALP-CR proton Interactions}\label{App:Axion_WF}

The observed $\gamma$-ray differential intensity along a given line-of-sight direction $\vec{n}$, tracing a cosmological density field $g(\vec{n},\chi)$, is conventionally parameterized as \cite{Fornengo:2013rga}
\begin{equation}\label{eq:gamma_intensity}
    I_{\gamma}(\vec{n}) = \int{\rm d}\chi \frac{g(\vec{n}, \chi)}{\langle g(\vec{n}, \chi) \rangle} {W}(\chi),
\end{equation}
where $\chi$ is the comoving distance, $W(\chi)$ encapsulates the line-of-sight physical contributions, and the brackets $\langle \cdots\rangle$ denote the ensemble average.

Fundamentally, the $\gamma$-ray intensity is determined by integrating the volume emissivity $P_{\gamma}$ over cosmic history \cite{Peacock:1999ye, Ando:2005xg}
\begin{equation}\label{eq:EI}
I_{\gamma}(\vec{n}, E_{\gamma}) = \frac{c}{4\pi E_{\gamma}} \int {\rm d} z \frac{P_{\gamma}([1+z]E_{\gamma}, \vec{n},z)}{H(z) (1+z)^4}e^{-\tau([1+z]E_{\gamma}, z)},
\end{equation}
where $H(z)$ is the Hubble parameter.

For the ALP-CR proton scattering process, the differential volume emissivity is defined by the target density, projectile flux, and scattering cross-section
\begin{align}
P_{\gamma}({E}_{\gamma}) &= {E}_{\gamma} n_{\rm DM} n_{\mathrm{p}} v_{\mathrm{p}} \frac{\mathrm{d}\sigma_{a + \mathrm{p}  \rightarrow \mathrm{p} + \gamma}}{\mathrm{d}E_{\gamma}} \nonumber \\
&= 4\pi {E}_{\gamma} n_{\rm DM}(\mathbf{x}|M) A_{\mathrm{p}}(M) f_{\mathrm{p}}(\mathbf{x}) \int_{E_{\mathrm{p}}^{\mathrm{min}}({E}_{\gamma})} \mathrm{d}E_{\mathrm{p}} f_{\mathrm{p}}(E_{\mathrm{p}}) \frac{\mathrm{d}\sigma_{a+\mathrm{p} \rightarrow \mathrm{p} + \gamma}}{\mathrm{d}E_{\gamma}}({E}_{\gamma}, E_{\mathrm{p}}, m_{a}), \label{eq:gamma_emissivity}
\end{align}
where $n_{\rm DM} = \rho_{\rm DM}/m_{a}$ is the ALP DM number density, and we have substituted the CR proton density relation
\begin{equation}
    \frac{\partial n_{\rm p}}{\partial E_{\rm p}} = \frac{4\pi}{v_{\rm p}}\frac{{\rm d}\Phi_{\rm p}}{{\rm d}E_{\rm p}}.
\end{equation}

The differential cross-section for the scattering is analytically given by \cite{Dent:2020qev}
\begin{align}
\frac{\mathrm{d}\sigma_{a + \mathrm{p} \rightarrow \mathrm{p} + \gamma}}{\mathrm{d}E_{\gamma}} &= \frac{1}{32\pi m_{a} |\vec{p}|^{2}} \left(\frac{1}{2}\sum_{\mathrm{spin}} |\mathcal{M}|^{2}\right) \nonumber \\
&= \frac{e^{2}g_{a\mathrm{p}}^{2}}{32\pi m_{a}|\vec{p}|^{2}} \Bigg[ \frac{4\Big(m_{\mathrm{p}}^{4}+3m_{a}^{2}m_{\mathrm{p}}^{2}-m_{\mathrm{p}}^{2}(2s+t) + (s-m_{a}^{2})(s+t)\Big)}{(s-m_{\mathrm{p}}^{2})(u-m_{\mathrm{p}}^{2})} \nonumber \\
&\quad - \frac{2\Big(m_{\mathrm{p}}^{4} - m_{\mathrm{p}}^{2}(2m_{a}^{2} + s + u) + su\Big)}{(u - m_{\mathrm{p}}^{2})^{2}} - \frac{2\Big(m_{\mathrm{p}}^{4} - m_{\mathrm{p}}^{2}(2m_{a}^{2}+s+u)+su\Big)}{(s - m_{\mathrm{p}}^{2})^{2}} \Bigg],
\end{align}
where $\mathcal{M}$ is the associated scattering amplitude, $|\vec{p}|^{2} = E_{\mathrm{p}}^{2} - m_{\mathrm{p}}^{2}$, and the Mandelstam variables are $s = m_{\mathrm{p}}^{2} + m_{a}^{2} + 2E_{\mathrm{p}}m_{a}$, $t = m_{a}^{2} - 2{E}_{\gamma}m_{a}$, and $u = 2m_{\mathrm{p}}^{2} + m_{a}^{2} - s - t$.

Using the relation $\mathrm{d}\chi = c\,\mathrm{d}z / H(z)$, we substitute Eq.~\eqref{eq:gamma_emissivity} into Eq.~\eqref{eq:EI} to obtain the line-of-sight integral
\begin{equation}
I_{\gamma}(E_{\gamma}) = \int \mathrm{d}\chi \frac{\rho_{\rm DM}(\mathbf{x}|z, M) A_{\mathrm{p}}(z, M) f_{\mathrm{p}}(\mathbf{x})}{m_{a}(1+z)^{3}} \int_{E_{\rm p}^{\rm min}([1+z]E_{\gamma})} \mathrm{d}E_{\mathrm{p}} f_{\mathrm{p}}(E_{\mathrm{p}}) \frac{\mathrm{d}\sigma_{a + \mathrm{p}  \rightarrow \mathrm{p} + \gamma}}{\mathrm{d}E_{\gamma}}([1+z]E_{\gamma}, E_{\mathrm{p}}, m_{a}) e^{-\tau([1+z]E_{\gamma}, z)}.
\end{equation}
Defining the spatial intensity field as $g_{a}(\mathbf{x}|z, M) = \rho_{\rm DM}(\mathbf{x}|z, M) A_{\mathrm{p}}(z, M) f_{\mathrm{p}}(\mathbf{x})$ and matching the integral to the definition in Eq.~\eqref{eq:gamma_intensity}, we directly extract the theoretical window function for ALP-CR proton interaction
\begin{equation}
W_{\mathrm{ap}}(E_{\gamma},\chi) = \frac{\langle g_{a}\rangle(z)}{m_{a}(1+z)^{3}} \int_{E_{\rm p}^{\rm min}([1+z]E_{\gamma})} \mathrm{d}E_{\mathrm{p}} f_{\mathrm{p}}(E_{\mathrm{p}}) \frac{\mathrm{d}\sigma_{a + \mathrm{p}  \rightarrow \mathrm{p} + \gamma}}{\mathrm{d}E_{\gamma}}([1+z]E_{\gamma}, E_{\mathrm{p}}, m_{a}) e^{-\tau([1+z]E_{\gamma}, z)},
\end{equation}
which concludes the derivation.

\clearpage
\twocolumngrid

\bibliographystyle{unsrt}
\bibliography{sample}

\end{document}